\documentstyle[11pt]{article}
\input psfig.tex
\begin{document}                                                                

\centerline{\bf Probing the Cosmic Large-Scale Structure with the} 
\vskip0.3cm

\centerline{\bf REFLEX Cluster Survey : Profile of an ESO Key Programme}
\vskip 0.5cm 

H. B\"ohringer$^{1}$, L. Guzzo$^2$, C.A. Collins$^3$,
D.M. Neumann$^4$, S. Schindler$^3$,\\
P. Schuecker$^{1}$,
R. Cruddace$^5$, S. DeGrandi$^2$, G. Chincarini$^2$, A.C. Edge$^6$,\\
H.T. MacGillivray$^7$, P. Shaver$^8$, G. Vettolani$^9$, W. Voges$^{1}$


\medskip
\noindent
{\it $^1$Max-Planck-Institut f\"ur extraterr. Physik, D-85740 Garching, Germany}\\
{\it $^2$Osservatorio Astronomico di Brera, Milano/Merate, Italy}\\
{\it $^3$Liverpool John-Moores University, Liverpool, U.K.}\\
{\it $^4$CEA Saclay, Service d'Astrophysique, Gif-sur-Ivette, France}\\
{\it $^5$Naval Research Laboratory, Washington, D.C., USA}\\
{\it $^6$Durham University, Durham, U.K.}\\
{\it $^7$Royal Observatory, Edinburgh, U.K.}\\
{\it $^8$European Southern Observatory, Garching, Germany}\\
{\it $^9$Istituto di Radioastronomia del CNR, Bologna, Italy}

\medskip

\section{Introduction}

To understand the formation of the visible structure in the 
Universe out of an initially almost homogeneous matter distribution
is one of the most fascinating quests of modern cosmology. 
The first step to this understanding is of course an assessment 
of the matter distribution in the present Universe on very
large scales extending over several hundred Mpc 
(for a Hubble constant of 50 km s$^{-1}$ Mpc$^{-1}$). 
Such large scales are
interesting for two major reasons: the present day structures on
this scale are directly comparable to the signature of the primordial 
structure detected in the microwave background by COBE with
similar co-moving sizes, and at these scales the observed density
fluctuations are just linear amplifications of the initial conditions
in the early Universe. While the study of the galaxy distribution has
already given us very interesting insights into the structure on scales of
a few hundred Mpc an alternative approach using the next larger
building blocks of the Universe, galaxy clusters, as probes for the
cosmic structure can give us a ready access to even larger scales
(see also Tadros et al. 1998 for the APM cluster survey in the optical).

X-ray astronomy has offered a unique tool to efficiently detect and
characterize galaxy clusters out to large distances. Originating in the
hot intracluster plasma that fills the gravitational potential well of
the clusters, the X-ray emission is an equally robust parameter for a
first estimate of the size and mass of clusters as the velocity
dispersion measured from the galaxy redshifts. But, while the X-ray 
luminosity can be readily detected for many clusters in an X-ray sky
survey, the collection of velocity dispersions requires very large and
time consuming redshift surveys. 

In the project described here we have embarked on a redshift survey 
of galaxy clusters detected in the ROSAT All-Sky Survey
(Tr\"umper 1993, Voges et al. 1996), the first
complete sky survey conducted with an imaging X-ray telescope.
In the frame of an ESO key programme 
(B\"ohringer 1994, Guzzo et al. 1995) we are seeking a definite 
identification of all possible cluster candidates found in the 
ROSAT Survey in the southern celestial hemisphere
above an X-ray flux limit chosen such to provide a reasonably 
homogeneous sensitivity coverage of the sky area. Within this 
survey programme, which we call the ROSAT ESO Flux Limited X-ray 
(REFLEX) Cluster Survey, we are investigating about 800 galaxy 
clusters and have obtained more than 300 new cluster redshifts.

The present article describes methods of the cluster detection
and identification as well as first preliminary results on
the statistics of the cluster population and a view on the
large-scale distribution of the clusters.
In a following paper we describe in more detail the
observing strategy and the spectroscopic results.

\section{Cluster Detection and Follow-Up Observations}

\begin{figure}[t]
\centerline{
\psfig{figure=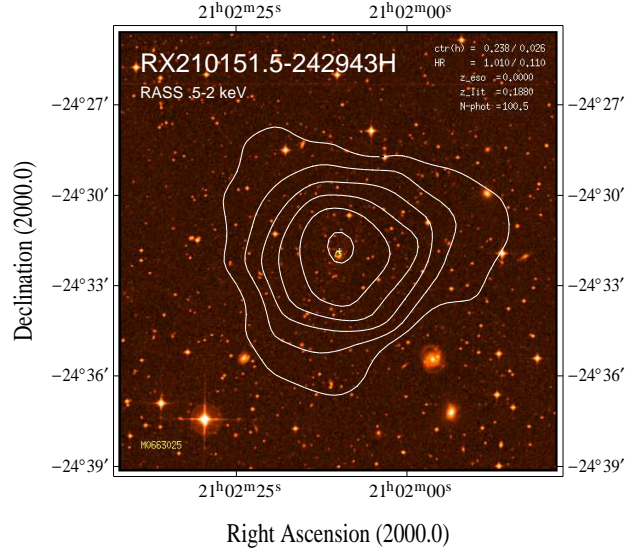,height=7.0cm}}
\caption{One of the galaxy clusters with a redshift
of $ z = 0.188$ not listed in the 
catalogue by Abell et al. (1989) found during the
identification of X-ray cluster sources from the ROSAT All-Sky Survey
in the frame of the ESO key programme. 
The figure has been produced from the STScI digital scans
of the UK-Schmidt IIIaJ plates with X-ray cotours from
the ROSAT All-Sky Survey images superposed. The contours are 
1.5,2,3,4,... sigma significance contours of the X-ray signal 
accounting properly for the photon noise and backgound.}
\end{figure}

The X-ray sky atlas constructed from the ROSAT All-Sky Survey Mission
with its more than 100,000 X-ray sources (Voges et al. 1996)
contains thousands of mostly unidentified galaxy clusters. 
Since the majority of these sources are characterized by few photons, 
only the brightest, well extended X-ray cluster sources are 
readily identified, while the
main part of the identifications has to be based on further
optical information. For the REFLEX Survey we made mainly
use of the COSMOS optical object catalogue 
(e.g. Heydon-Dumbleton et al. 1989) produced from 
the digital scans of the photographic UK Schmidt Survey ( providing
star/galaxy separation for the sky objects with high completeness
down to $b_j = 20.5$ mag.). The cluster candidates are found
as overdensities in the galaxy density distribution at the
position of the X-ray sources (see B\"ohringer et al. in preparation).    
Not all the sources flagged by this technique are true galaxy clusters,
however.
The price paid for aiming at a high completeness in the final 
catalogue is a low detection threshold in the galaxy density 
leading to a contamination of 
the candidate list by more than 30\% non-cluster sources. This large
contamination can be reduced to about one third by a direct inspection 
of the photographic plates, the detailed X-ray properties, and 
using all the available literature information. 

The subsequent work is the observational task of the ESO key programme
comprising a total of 90 observing nights distributed evenly among 
the 1.5m, 2.2m, and 3.6m telescopes at La Silla.
These follow-up observations have a two-fold purpose: 
we search for a definite identification for the unknown objects  
and collect redshifts for all clusters for which this parameter is not known.
To make optimal use of the ESO telescopes we observe nearby, poor clusters 
and groups in single-slit mode with the smaller telescopes and use the
3.6m telescope with EFOSC in multi-slit operation to get multiple
galaxy spectra
of dense, rich clusters. Several coincident redshifts strongly support
the cluster identification and the spectroscopy of central dominant 
early type galaxies at the X-ray maximum plays a particularly important 
role in this study.

The observing programme will be completed in fall of this year.
Having observed the clusters with higher X-ray flux with highest
priority, we can construct a first catalogue of clusters
for the brighter part of the sample down to an X-ray flux limit of
$3 \cdot 10^{-12}$ erg s$^{-1}$ cm$^{-2}$ (in the ROSAT band 0.1 - 2.4 keV)
comprising 475 objects.
53\% of these clusters can be found in the main catalogues of Abell (1958)
and Abell, Corwin, \& Olowin (1989) and further 10\% in the supplementary
list. Most of the remaining clusters are previously
unknown objects. This result highlights the importance of the selection
process in the construction of the sample: a significant fraction
of the X-ray bright and massive clusters would have been missed had the
X-ray clusters only been identified on the basis of existing catalogues. 
The fraction of known Abell clusters decreases further if one goes
to lower X-ray flux limits e.g. in the extended REFLEX sample.
We should also note here, however, that one part of the objects missing
in Abell's compilation are nearby, X-ray bright groups, which are 
not rich enough in their galaxy content to fulfill Abell's criteria.

The high completeness of this sample is demonstrated
by a counter-test, in which we searched for X-ray emission at the position
of all Abell and ACO clusters in the ROSAT Survey independent of a previous
detection of these sources by the survey source identification algorithm.
Only 5 clusters ($\sim 1\%$) were found to have been missed by the cluster 
search based on the COSMOS data. 
Fig. 1 gives an example of a non-Abell clusters found
at a redshift of $z =  0.188$. A fraction of the
X-ray sources in the cluster candidate list are found to have non-cluster
counterparts in the follow-up observations including AGN in clusters where the
AGN provide the main source of the X-ray emission.   
 
A few spectacular discoveries were made in the course of this programme
including the most X-ray luminous cluster discovered so far, RXJ1347-1145
(Schindler et al. 1995,1996) and 
clusters featuring bright gravitational arcs, e.g. S295 
(Edge et al. 1994).

\section{Properties of the X-ray Clusters of Galaxies}

\begin{figure}[t]
\psfig{figure=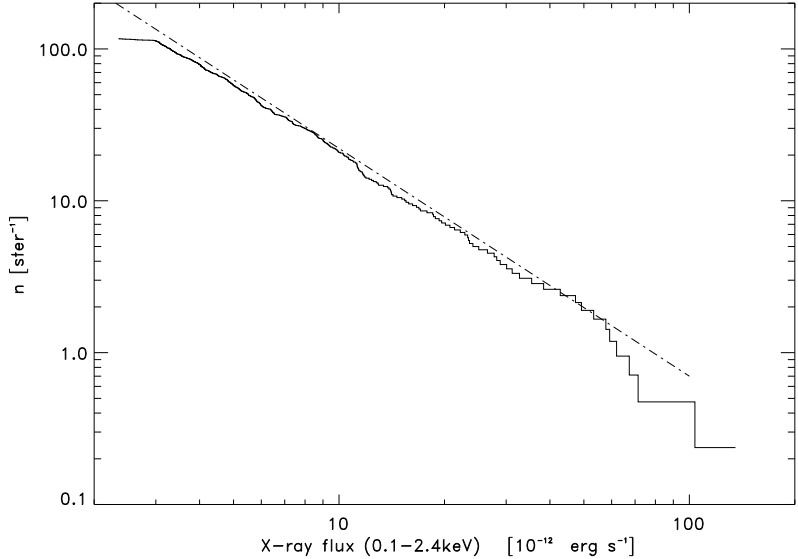,height=8cm}
\caption{Cumulative number counts of X-ray clusters as a function
of the limiting flux for the REFLEX sample.}
\end{figure}

In the following we are reporting results based on the X-ray bright sample
of 475 galaxy clusters for which 413 cluster redshifts have been obtained
and reduced so far. A plot of the number counts of the cluster population
as a function of the limiting X-ray flux is shown in Fig. 2. The logarithmic
graph has a slope that is in perfect agreement with an Euclidian 
slope of $-3/2$.
This slope is easily explained by the fact that the majority of the clusters
is not very distant - the peak in the redshift histogramme is at $z \sim 0.06$ -
and thus band-corrections and evolutionary corrections are not important. 
Fig. 3 shows the distribution in redshift and X-ray luminosity. While most of
the clusters detected are not very distant (median redshift 
is $z=0.085$), a few
very luminous clusters are found out to redshifts of $z = 0.3$ with one
outstanding object at a redshift of $z = 0.45$, the most luminous 
cluster mentioned above. One also notes clearly the break at a redshift
of $z = 0.3$, which is caused by the limited depth of the optical plate material
used for the cluster pre-identification. Thus we can expect that there are 
more, in fact very interesting luminous clusters in the X-ray source list
for this flux limit with redshifts between $z = 0.3$ to $0.5$. 
But a more expensive imaging search programme would be necessary to find these
objects.

Another very interesting part of this X-ray cluster population is found among
the nearby, low X-ray luminosity objects.  These are groups of galaxies
dominated by very giant cD galaxies which are often found to be two or
three magnitudes brighter than the next brightest galaxy in the 
system. A CCD image of such a group is shown in Fig. 4.

\begin{figure}[t]
\psfig{figure=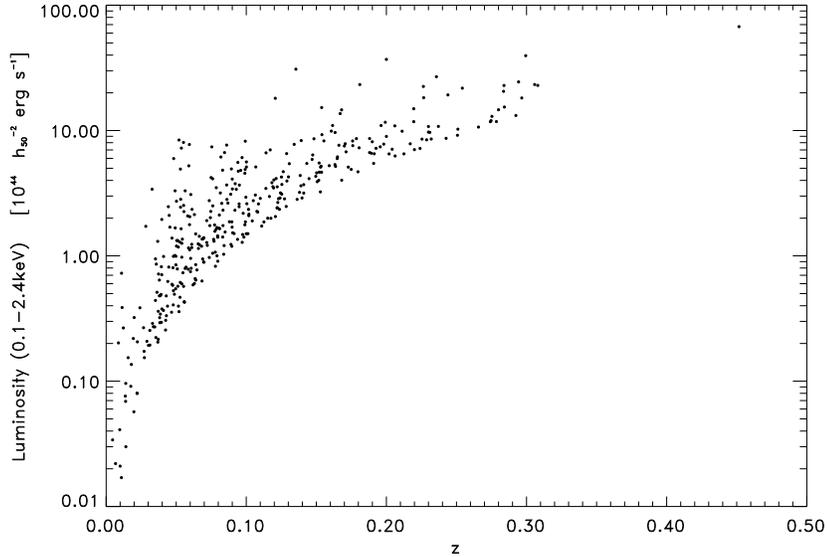,height=8cm}
\caption{Distribution of the X-ray luminosities as a function of redshift
for the galaxy clusters of the REFLEX sample.}
\end{figure}

\begin{figure}[t]
\centerline{
\psfig{figure=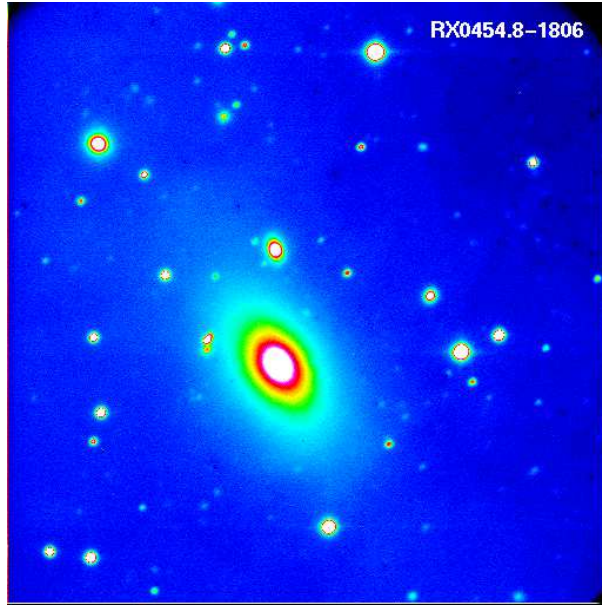,height=8cm}}
\caption{Galaxy group with a very dominant central cD galaxy
at a redshift of $z = 0.0313$ in the REFLEX sample. 
The CCD image is a short routine exposure (without filter)
to define the MOS slit mask taken with EFOSC1 at 
the 3.6m telescope. The scale of the image is about 3.5 by 3.5 arcmin.}

\end{figure}

The most straightforwardly obtained and 
important distribution function of the cluster sample is the X-ray
luminosity function. This function is most closely related to the mass
function of the clusters which is used as an important calibrator
of the amplitude of the density fluctuation power spectrum of the 
Universe (e.g. White et al. 1993). A preliminary version of the X-ray
luminosity function of the sample is shown in Fig. 5. Note that 
this function was derived for the sample when ($ \sim 80 \%$) of
the redshifts had been obtained.
But in spite of this incompleteness the luminosity
function already recovers the densities reached in previous surveys
(e.g. Ebeling et al. 1997, DeGrandi 1996) to which the present result is
compared in the figure. The high quality of the present sample is also 
demonstrated by the fact that the slope of the number count function
of Fig. 2 is Euclidian while the previous surveys showed a significantly
lower value for the logarithmic slope of $\sim -1.35$.

\begin{figure}[t]
\psfig{figure=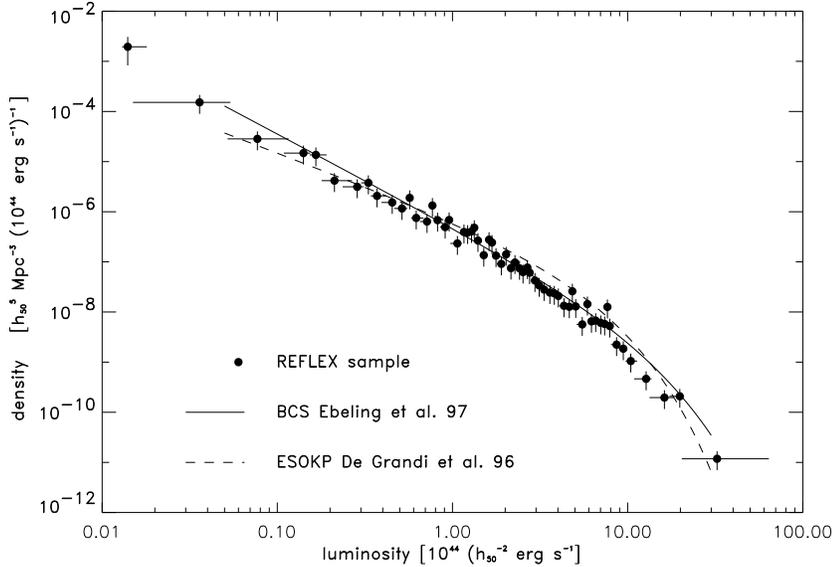,height=8cm}
\caption{Differential X-ray luminosity function for the REFLEX sample
(points with error bars; for details see B\"ohringer et al. 
in preparation) compared to the previous results by
DeGrandi (1996) and Ebeling et al. (1997). A value of $H_0 = 50$
km s$^{-1}$ Mpc$^{-1}$ is assumed for the scaling. }
\end{figure}

As pointed out in the introduction, the most exciting 
aim of this programme is to
assess the large-scale structure of the galaxy cluster and matter distribution
in the Universe. The currently most popular and important statistical function
for the characterization of the large-scale structure is the density 
fluctuation power spectrum. A first preliminary result for this function
obtained from 188 cluster of our sample in a box of 400 Mpc is shown
in Fig. 6 (for details see Schuecker et al. in preparation). The function
is featuring an interesting maximum at a scale of about 100 - 150 
$h_{100}^{-1}$ Mpc.
The location of this maximum is related to the size of the horizon of the
Universe at an epoch when the energy density of matter and radiation was
equal and it is therefore a very important calibration point for the
theoretical modeling of the evolution of our Universe. Note that the
construction of the power spectrum shown in Fig. 6 is based on a 
luminosity (mass) selection of the clusters which varies with 
redshift. Therefore the quantitative interpretation of the this
result is not straightforward and more work is needed to relate
this function to the power spectrum of the matter density fluctuation
in the Universe.  

\begin{figure}[t]
\centerline{
\psfig{figure=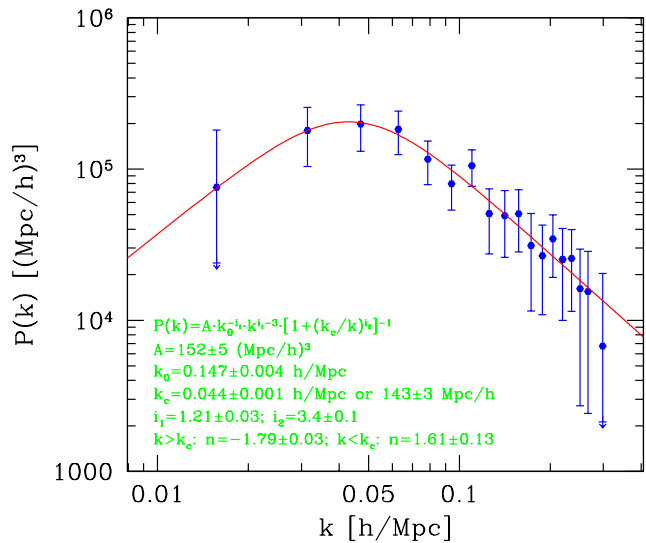,height=9cm}}
\caption{Power spectrum of the density distribution of the galaxy clusters
in the REFLEX sample. For the analysis only 188 clusters in a box of
400 Mpc side length was used (see Schuecker et al. in preparation).
A value of $H_0 = 100$ km s$^{-1}$ Mpc$^{-1}$ is assumed for the scaling.
The solid line is a parametric fit to the data not considering a particular
cosmological model.}
\end{figure}

From the power spectrum shown we can also derive another statistical 
function which is more illustrative to non-experts, the rms-fluctuation
level on different scales as provided by a filtering of the cluster density
fluctuation field by a simple Gaussian smoothing filter. This function
is shown in Fig. 7 where we note that at the scale of the maximum of the
power spectrum (a scale of $\sim 100~h_{100}^{-1}$ Mpc) the cluster density 
varies typically by about 10\%, while at the largest scales 
($400~h_{100}^{-1}$ Mpc) a 
fluctuation level of the order of 1\% is indicated. This sets a high
standard of requirements on the cluster survey, and while we are confident
that the present sample contains no systematic biases on the 10\%
level, further tests and simulations are needed to explore the reliability
of the results on the largest scales. 

\begin{figure}[t]
\psfig{figure=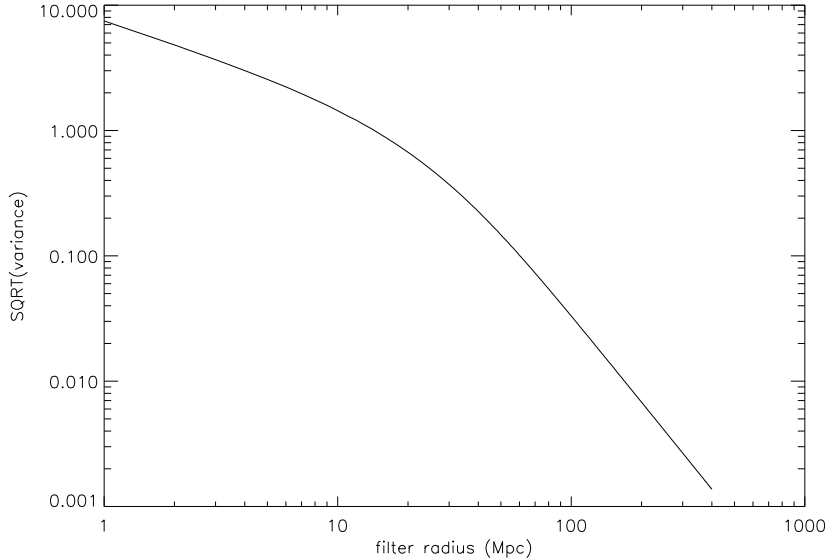,height=8cm}
\caption{Square root of the variance of the fluctuations
in the cluster density distribution as a function of scale
(obtained with Gaussian filtering).
A value of $H_0 = 100$ km s$^{-1}$ Mpc$^{-1}$ is assumed for the scaling.}
\end{figure}

\section{Conclusions}

It is obvious from the above arguments that studies of the large-scale
structure using clusters as probes require great care in the 
preparation of the sample of test objects. The first results
of the REFLEX Survey presented here and some further tests
that we have conducted already demonstrate the high quality of the
present sample. This is the result of a very homogeneous and highly
controlled selection process used to construct the sample.
The challenge of the selection work in this survey is to 
combine the complementary information from X-ray and optical wavelength
in the most homogeneous way. Contrary to several earlier studies
that we have conducted (Pierre et al. 1994, Romer et al. 1994, Ebeling et al.
1997, De Grandi 1996) in which the identification process was
optimized to find as many clusters as possible by using all available
sources of information, we have now achieved a highly complete 
cluster selection by just combining the ROSAT All-Sky Survey data
and the COSMOS optical data base in a homogeneous way, completely
controlled by automated algorithms. Additional information is
only used in the final identification but does
not influence the selection. This is a very important achievement
in this survey work. 

The data presented are still not fully complete in redshifts.
But with data already obtained
in January and September 1998 we can practically
complete this data set (to 96\%). An extended
sample of REFLEX clusters down to a flux limit of $2\cdot 10^{-12}$ erg
s$^{-1}$ cm$^{-2}$ is already prepared and redshifts are available for
more than 70\% of the objects. This extended set of about 750 galaxy
clusters will help very
much to tighten the constraints for the power spectrum and extend it
to larger scales. It will further enable us to investigate the cluster
correlation function - in particular the X-ray luminosity 
dependence of the
clustering amplitude, which is an issue not yet resolved.  
Finally, a complementary ROSAT Survey cluster identification
programme is being conducted in the Northern Sky in a collaboration of
the Max-Planck Institut f\"ur extraterrestrische Physik and
J. Huchra, R. Giacconi, P. Rosati and B. McLean which will soon
reach a similar depth and provide an all-sky view on the X-ray cluster
distribution.

\end {document}